\documentclass[reprint,twocolumn,superscriptaddress,prl]{revtex4-1}

\usepackage{amsmath,amssymb}
\usepackage{graphicx}
\usepackage{dcolumn}
\usepackage{bm}
\usepackage{xr}

\newcommand{\Dc}{D_\mathrm{c}}
\newcommand{\Dm}{D_\mathrm{m}}
\newcommand{\etaM}{\eta_{\mathrm{M}}}
\newcommand{\etaS}{\eta_\mathrm{S}}
\newcommand{\etaT}{\eta_\mathrm{T}}
 
\newcommand{\Wx}{\dot{W}_X}
\newcommand{\Wy}{\dot{W}_Y}
\newcommand{\Qx}{\dot{Q}_X}
\newcommand{\Qy}{\dot{Q}_Y}
\newcommand{\Wxy}{\dot{W}_{X\to Y}}
\newcommand{\Wyx}{\dot{W}_{Y\to X}}
\newcommand{\Ix}{\dot{I}_X}
\newcommand{\Iy}{\dot{I}_Y}
\newcommand{\Sigmax}{\dot{\Sigma}_X}
\newcommand{\Sigmay}{\dot{\Sigma}_Y}
\newcommand{\etax}{\eta_X}
\newcommand{\etay}{\eta_Y}

\begin{document}

\title{Inferring Subsystem Efficiencies in Bipartite Molecular Machines}

\author{Matthew P.\ Leighton}
\email{matthew\_leighton@sfu.ca}
\author{David A.\ Sivak}%
\email{dsivak@sfu.ca}
\affiliation{Department of Physics, Simon Fraser University, Burnaby, British Columbia, V5A 1S6, Canada.}%

\date{\today}

\begin{abstract}
Molecular machines composed of coupled subsystems transduce free energy between different external reservoirs, in the process internally transducing energy and information. While subsystem efficiencies of these molecular machines have been measured in isolation, less is known about how they behave in their natural setting when coupled together and acting in concert. Here we derive upper and lower bounds on the subsystem efficiencies of a bipartite molecular machine. We demonstrate their utility by estimating the efficiencies of the $\mathrm{F}_\mathrm{o}$ and $\mathrm{F}_1$ subunits of ATP synthase and that of kinesin pulling a diffusive cargo.
\end{abstract}


\maketitle

Molecular machines are integral to the functioning of all living organisms, accomplishing tasks within cells by transducing energy between different forms~\cite{brown2019theory}. A molecular machine can also transduce energy within itself, between internally coupled components~\cite{large2021free,mcgrath2017biochemical}. Two paradigmatic examples are $\mathrm{F}_\mathrm{o}\mathrm{F}_1-$ATP synthase, which converts electrochemical energy from a transmembrane proton gradient into the synthesis of ATP molecules via free energy transduction between the $\mathrm{F}_\mathrm{o}$ and $\mathrm{F}_1$ subsystems~\cite{oster1999atp}, and transport motors like kinesin that transduce chemical energy in the form of ATP into mechanical work pulling molecular cargo against viscous friction~\cite{woehlke2000walking}.

In addition to biological molecular machines like the above examples, it is now also possible to design \textit{de novo} and assemble two-component molecular machines~\cite{wilson2016autonomous,courbet2022computational}. To facilitate future design of synthetic molecular machines, it is critical to understand the functionality and performance of existing molecular machines optimized by evolutionary forces, and engineering principles governing particularly effective machines~\cite{brown2017toward}.

The theory of stochastic thermodynamics~\cite{seifert2012stochastic} facilitates these efforts, quantifying the energetics of stochastic systems and enabling inference of thermodynamic properties from observations of a system's dynamical behavior~\cite{seifert2019stochastic}. Autonomous two-component systems like molecular machines can exchange energy~\cite{li2016reaction}, free energy~\cite{large2021free}, and information~\cite{horowitz2014thermodynamics,hartich2014stochastic,mcgrath2017biochemical}. Using this framework, specific models of bipartite molecular machines such as $\mathrm{F}_\mathrm{o}\mathrm{F}_1-$ATP synthase~\cite{okazaki2015elasticity,ai2017torque,lathouwers2020nonequilibrium,lathouwers2022internal}, transport motors pulling cargo~\cite{zimmermann2015effective,brown2019pulling,leighton2022performance}, and even synthetic molecular motors~\cite{amano2022insights} have been studied to understand various performance trade-offs that shape their design and behavior.

Subsystems of bipartite molecular machines, like the $\mathrm{F}_1$ subunit~\cite{toyabe2011thermodynamic}, have been studied in isolation to determine their efficiency. Less, however, is known about how these subsystems perform when coupled together, as when performing their functions inside of biological organisms. For example, while experiments that measure motor efficiency typically apply a constant force, modeling efforts have shown that transport motors perform differently when pulling a diffusive cargo~\cite{zimmermann2015effective,brown2019pulling}. Understanding molecular machines thus requires estimates of subsystem efficiencies within bipartite machines, in addition to their efficiencies in isolation.

In this work we study the stochastic thermodynamics of autonomous bipartite molecular machines, detailing a new method to derive upper and lower bounds on the thermodynamic efficiencies of bipartite subsystems from any bounds on subsystem entropy production rates. As an example, we apply the recently proven Jensen lower bounds~\cite{leighton2022dynamic}, which do not depend on detailed internal interactions, making them easy to compute even from limited data. We illustrate the utility of these bounds using experimental measurements to infer the efficiencies of $\mathrm{F}_\mathrm{o}$ and $\mathrm{F}_1$ when coupled together, as well as the efficiency of a kinesin motor while pulling a diffusive vesicular cargo. Ultimately our method allows for measurements of the efficiencies of subsystems in their natural settings, something inaccessible when studying them in isolation.

\emph{Stochastic thermodynamics of bipartite systems}.---Consider an autonomous molecular machine with two continuous degrees of freedom $x$ and $y$ denoting the coordinates of the $X$ and $Y$ subsystems. Each coordinate evolves according to an overdamped Langevin equation:
\begin{subequations}\label{LangevinEqs}
\begin{align}
\dot{x} & = \beta D_X\left[ f_X(x,y) - \frac{\partial }{\partial x}V(x,y)\right] + \sqrt{2D_X}\, \xi_X(t) ,\\
\dot{y} & = \beta D_Y\left[f_Y(x,y) - \frac{\partial }{\partial y}V(x,y)\right]+ \sqrt{2D_Y}\, \xi_Y(t) \ .
\end{align}
\end{subequations}
The potential $V(x,y)$ captures interactions between $X$ and $Y$ as well as any subsystem-specific energy features, $f_X(x,y)$ and $f_Y(x,y)$ are nonconservative forces, and $\xi_X(t)$ and $\xi_Y(t)$ denote Gaussian white noises. We assume the dynamics are bipartite, meaning that $\xi_X(t)$ and $\xi_Y(t)$ are independent. 

The subsystem diffusion coefficients are $D_X$ and $D_Y$, which we assume are related to the friction coefficients $\zeta_X$ and $\zeta_Y$ by the fluctuation-dissipation relation $\beta D_X\zeta_X = 1 = \beta D_Y\zeta_Y$ as is commonly done in stochastic thermodynamics~\cite{seifert2012stochastic}. $\beta\equiv\left(k_\mathrm{B}T\right)^{-1}$ is the inverse temperature. These are ``bare'' diffusion and friction coefficients~\cite{felderhof1978diffusion}, rather than ``effective'' coefficients that convolve the influences of the potential and nonequilibrium driving forces.

\begin{figure}[h] 
\includegraphics[width=\columnwidth]{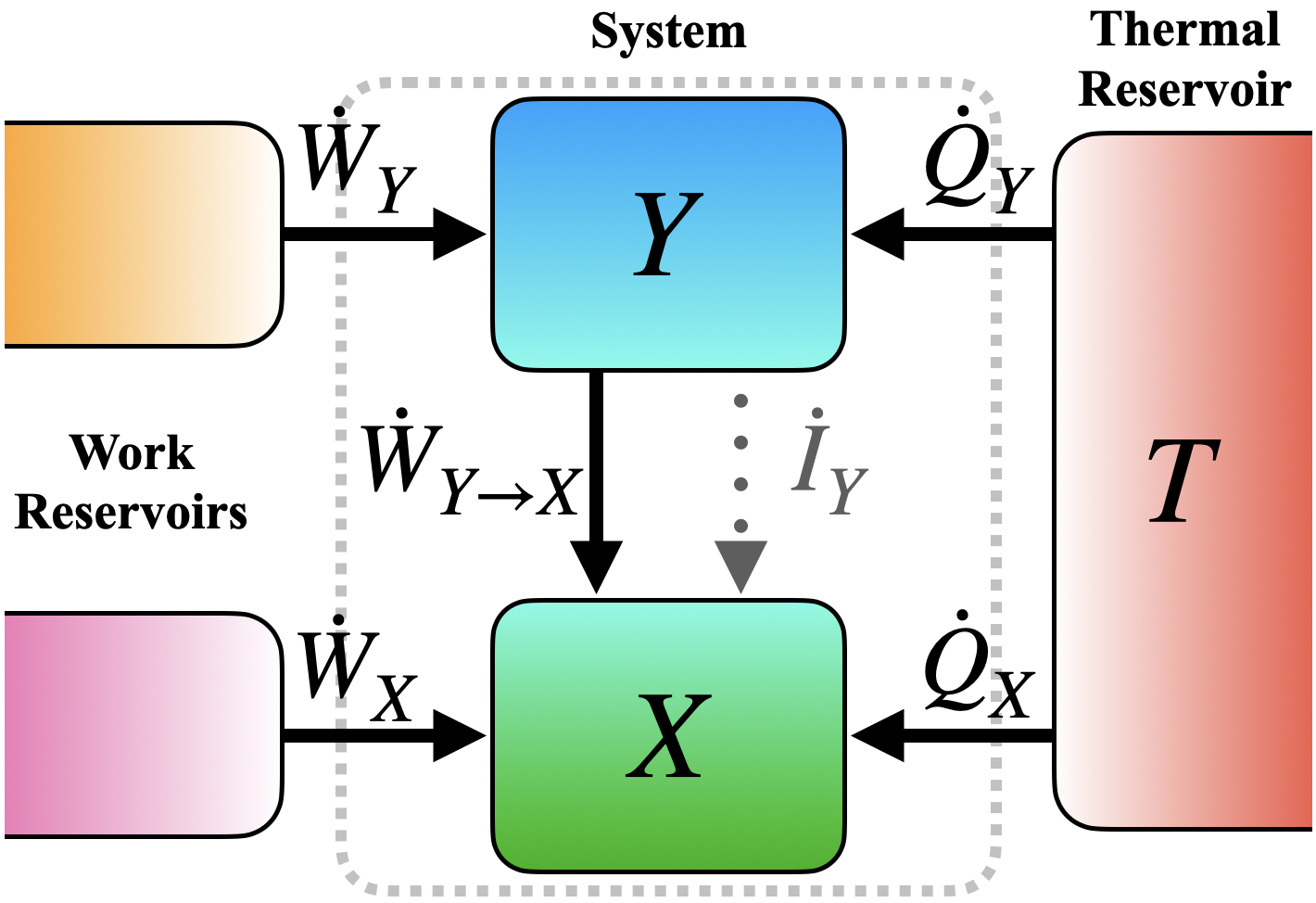}
\caption{\label{fig:Energy_Diagram} Steady-state energy and information flows for a bipartite molecular machine. The $X$ and $Y$ subsystems can exchange work and heat with external reservoirs, and energy and information with each other. Arrows indicate positive flows under our sign convention.}
\end{figure}

We now restrict our attention to the nonequilibrium steady state. Each subsystem ($X$ and $Y$) exchanges work and heat with external reservoirs (Fig.~\ref{fig:Energy_Diagram}). The average rate of external work into the $Y$ subsystem is
\begin{equation}
\Wy \equiv \left\langle f_Y(x,y)\circ \dot{y}\right\rangle,
\end{equation}
while the average rate of heat into $Y$ from the environment is
\begin{equation}
\Qy \equiv \left\langle \left[\frac{\partial V}{\partial y} - f_Y(x,y)\right]\circ \dot{y}\right\rangle.
\end{equation}
Here angle brackets denote ensemble averages, and the symbol ``$\circ$" indicates the Stratonovich product~\cite{seifert2012stochastic}. Finally, treating $y$ as an external control parameter driving $X$, the transduced work from $Y$ to $X$ is defined analogously to Ref.~\cite{sekimoto1998langevin} as~\cite{large2021free,ehrich2022energy}
\begin{equation}
\Wyx \equiv \left\langle \frac{\partial V}{\partial y}\circ \dot{y}\right\rangle.
\end{equation}
The interpretation of $\Wyx$ as a work rate relies on the steady-state assumption made above; outside of steady state, this quantity more generally quantifies the change in system energy due to the dynamics of $Y$~\cite{ehrich2022energy}. Analogous quantities $\Wx$, $\Qx$, and $\Wxy$ can likewise be defined for energy flows into and out of the $X$ subsystem.

Each subsystem obeys a first law describing local energy conservation:
\begin{subequations}\label{firstlaws}
\begin{align}\label{xfirstlaw}
\Wx + \Qx & = \Wxy,\\\label{yfirstlaw}
\Wy + \Qy & = \Wyx.
\end{align}
\end{subequations}
Likewise, each subsystem satisfies a subsystem-specific second law~\cite{horowitz2014thermodynamics}:
\begin{subequations}\label{secondlaws}
\begin{align}\label{xsecondlaw}
\Sigmax & = - \beta\Qx - \Ix\geq 0,\\\label{ysecondlaw}
\Sigmay & = - \beta\Qy - \Iy\geq 0.
\end{align}
\end{subequations}
Here $\Sigmax$ and $\Sigmay$ are the mean dimensionless entropy production rates of the $X$ and $Y$ subsystems, and
\begin{equation}\label{infoflowdef}
\Iy \equiv \left\langle \frac{\partial}{\partial y} \ln p_{X|Y}(x|y) \circ \dot{y}\right\rangle 
\end{equation}
is the information flow due to $Y$, quantifying the rate at which the dynamics of $Y$ increase the mutual information between $X$ and $Y$. $p_{X|Y}(x|y)$ is the conditional distribution of $X$ given $Y$. An analogous definition holds for $\Ix$. The definition \eqref{infoflowdef} in the Langevin formulation is equivalent to that given for Fokker-Planck dynamics in Ref.~\cite{horowitz2015multipartite}. We assume here that the system is only weakly coupled to its environment, and therefore there are no information flows between the subsystems and reservoirs.

At steady state, the transduced works and information flows satisfy $\Ix + \Iy=0$ and $\Wxy + \Wyx=0$. Eqs.~\eqref{firstlaws} and \eqref{secondlaws} thus combine to yield two inequalities constraining the transduced capacity $\beta\Wyx + \Iy$~\cite{lathouwers2022internal}:
\begin{subequations}\label{transducedcapacitybounds}
\begin{align}
\Sigmax & =  \beta\Wx + \beta\Wyx + \Iy \geq 0,\label{xsecondlaw}\\
\Sigmay & = \beta\Wy - \beta\Wyx - \Iy\geq 0.
\end{align}
\end{subequations}

\emph{Efficiency measures}.---Natural definitions of efficiency depend on the direction that energy flows through the system at steady state. Without loss of generality, let $Y$ be the ``upstream'' subsystem, such that $\Wy> 0$. We restrict attention to functional machines that output work ($\Wx\leq 0$); then by Eq.~\eqref{xsecondlaw}, $Y$ drives $X$ with non-negative transduced capacity ($\Wyx+\Iy \geq 0$).

The simplest measure of efficiency is the global thermodynamic efficiency $\etaT$$\equiv-\Wx/\Wy$, the ratio of the output work and the input work. By the global second law $0\leq\dot{\Sigma}=\Sigmax+\Sigmay$, the efficiency satisfies $\etaT\leq 1$. 

With the framework of Eqs.~\eqref{firstlaws} and \eqref{secondlaws}, the subsystems $X$ and $Y$ are thermodynamic systems in their own right, each satisfying local first and second laws. They thus each have their own thermodynamic efficiency:
\begin{subequations}\label{efficiencydefs}
\begin{align}
\etay &\equiv \frac{\beta\Wyx+\Iy}{\beta\Wy},\label{etaydef}\\
\etax &\equiv \frac{-\beta\Wx}{\beta\Wyx+\Iy}.\label{etaxdef}
\end{align}
\end{subequations}
Introduced in Ref.~\cite{barato2017thermodynamic} and later studied in Ref.~\cite{amano2022insights}, the subsystem efficiency $\etay$ quantifies the efficiency with which $Y$ transduces input work into available free energy for subsystem $X$, while $\etax$ quantifies how efficiently $X$ converts that free energy into output work. Their product is the global thermodynamic efficiency, $\etaT = \etay\etax$. These efficiencies are well-defined so long as $\Wy$ and $\beta\Wyx + \Iy$ are both strictly positive, with $0\leq \etax\leq 1$ and $0\leq \etay\leq 1$ then following from Eqs.~\eqref{transducedcapacitybounds}. Note that these efficiencies (Eqs.~\eqref{etaydef} and \eqref{etaxdef}) differ from the subsystem efficiency measures defined in Ref.~\cite{horowitz2014thermodynamics}, which track the efficiency of information usage rather than the efficiency of free-energy transduction.

Finally, when $\Wx=0$ and thus $\etaT=0$, the system may still perform useful work moving the $X$ subsystem against viscous friction. This is the case, for example, when $Y$ is a transport motor and $X$ a diffusive molecular cargo. In this case, an alternative measure of efficiency is the Stokes efficiency~\cite{wang2002stokes}
\begin{equation}\label{StokesEfficiency}
\etaS \equiv \frac{\zeta_X\langle \dot{x}\rangle^2}{\Wx + \Wy}\ ,
\end{equation}
the ratio of the work that would be done in moving $X$ at constant velocity $\langle \dot{x} \rangle$ against viscous friction and the external work into the system.

\emph{Bounds on subsystem efficiencies}.---Equations~\eqref{transducedcapacitybounds} provide two equalities for the transduced capacity,
\begin{equation}\label{eq:TCeqs}
\Sigmax - \beta\Wx = \beta\Wyx + \Iy = \beta\Wy - \Sigmay.
\end{equation}
Applying to Eq.~\eqref{eq:TCeqs} any lower bounds $\Sigmax^\mathrm{LB}\leq \Sigmax$ and $\Sigmay^\mathrm{LB}\leq \Sigmay$ on the subsystem entropy production rates yields upper and lower bounds on the transduced capacity:
\begin{equation}\label{eq:TCineqs}
\Sigmax^\mathrm{LB}- \beta\Wx \leq \beta\Wyx + \Iy \leq \beta\Wy - \Sigmay^\mathrm{LB}.
\end{equation}

Dividing Eq.~\eqref{eq:TCineqs} by $\Wy$ yields upper and lower bounds on $Y$'s efficiency:
\begin{equation}\label{eta_y_ineq_general} 
\etaT\left(1 + \frac{\Sigmax^\mathrm{LB}}{-\beta\Wx}\right)\leq \etay \leq 1 - \frac{\Sigmay^\mathrm{LB}}{\beta\Wy}.
\end{equation}
Likewise, multiplying the reciprocal of Eq.~\ref{eq:TCineqs} by $-\Wx$ yields upper and lower bounds on $X$'s efficiency:
\begin{equation}\label{eta_x_ineq_general} 
\etaT\left( 1 - \frac{\Sigmay^\mathrm{LB}}{\beta\Wy}\right)^{-1}  \leq \etax \leq \left( 1 + \frac{\Sigmax^\mathrm{LB}}{-\beta\Wx}\right)^{-1}.
\end{equation}
The two inequalities~\eqref{eta_y_ineq_general} and \eqref{eta_x_ineq_general} constitute the most general form of our main result, providing a recipe to derive bounds on subsystem efficiencies using the interactions of subsystems with their environments and lower bounds on their entropy production rates. These inequalities are valid for any lower bounds $\Sigmax^\mathrm{LB}$ and $\Sigmay^\mathrm{LB}$, and are also valid for discrete degrees of freedom.

Inserting the subsystem second laws $\Sigmax^\mathrm{LB}=0 = \Sigmay^\mathrm{LB}$ into Eqs.~\eqref{eta_y_ineq_general} and \eqref{eta_x_ineq_general} yields $\etaT \leq \eta_{X/Y}\leq 1$. Beyond the second law, however, the recently derived Jensen bound~\cite{leighton2022dynamic} gives tighter lower bounds for the overdamped bipartite Langevin dynamics considered here:
\begin{subequations}\label{JensenBounds}
\begin{align}
\beta\zeta_X\langle \dot{x}\rangle^2 & \leq \Sigmax,\\
\beta\zeta_Y\langle \dot{y}\rangle^2 & \leq \Sigmay.
\end{align}
\end{subequations}
$\langle \dot{x}\rangle$ and $\langle \dot{y}\rangle$ are the steady-state average rates of change of the coordinates $X$ and $Y$. Inserting these Jensen bounds into Eqs.~\eqref{eta_y_ineq_general} and \eqref{eta_x_ineq_general} gives 
\begin{subequations}\label{Jensen_ineqs}
\begin{equation}\label{eta_y_ineq} 
\etaT\left(1 + \frac{\zeta_X\langle\dot{x}\rangle^2}{-\Wx}\right)\leq \etay \leq 1 - \frac{\zeta_Y\langle\dot{y}\rangle^2}{\Wy},
\end{equation}
\begin{equation}\label{eta_x_ineq} 
\etaT\left( 1 - \frac{\zeta_Y\langle\dot{y}\rangle^2}{\Wy}\right)^{-1}  \leq \etax \leq \left( 1 + \frac{\zeta_X\langle\dot{x}\rangle^2}{-\Wx}\right)^{-1}.
\end{equation}
\end{subequations}
This is a specific, immediately applicable version of our main result. Equations~\eqref{Jensen_ineqs} bound internal energetic flows through subsystems, in terms of the experimentally accessible quantities $\zeta_X$, $\zeta_Y$, $\langle\dot{x}\rangle$, $\langle\dot{y}\rangle$, $\Wx$, and $\Wy$ (recall that $\etaT = -\Wx/\Wy$). These quantities solely depend on and characterize the interactions of the two subsystems with their environments; applying the bounds Eqs.~\eqref{eta_y_ineq} and \eqref{eta_x_ineq} does not require any knowledge of the details of the coupling between subsystems.

Molecular machines that transduce free energy into directed motion rather than into stored free energy will often produce no output work ($\Wx=0$, and thus also $\etax=0=\etaT$). It is then desirable to reformulate Eq.~\eqref{eta_y_ineq} in a way that incorporates the Stokes efficiency and does not include division by $\Wx$. Substituting the definition of $\etaT$, taking $\Wx=0$, and identifying $\etaS$~\eqref{StokesEfficiency}, Eq.~\eqref{eta_y_ineq} simplifies significantly to
\begin{equation}\label{finalmotorcargobound}
\etaS \leq \etay\leq 1 - \frac{\zeta_Y\langle \dot{y}\rangle}{\zeta_X\langle \dot{x}\rangle} \, \etaS.
\end{equation}

\emph{Subsystem efficiencies in ATP synthase}.---We now apply Eqs.~\eqref{eta_y_ineq} and \eqref{eta_x_ineq} to the molecular machine ATP synthase. The two coordinates $Y$ and $X$ correspond roughly to the rotational states of the c-ring inside the $\mathrm{F}_\mathrm{o}$ subsystem and the $\gamma$-shaft inside the $\mathrm{F}_1$ subsystem, respectively~\cite{lathouwers2020nonequilibrium}.

Lacking experimental data on ATP synthesis and proton translocation rates, we assume that the $\mathrm{F}_\mathrm{o}$ and $\mathrm{F}_1$ subsystems tightly couple~\cite{soga2017perfect} rotary motion with proton translocation~\cite{marciniak2022determinants} and ATP synthesis~\cite{toyabe2011thermodynamic}, respectively. The external work rates are then $\dot{W}_\mathrm{o} = \Delta\mu_{\mathrm{H}^+} \langle J_\mathrm{o}\rangle$ and $\dot{W}_1 = \Delta\mu_{\mathrm{ATP}}\langle J_1\rangle$, for the two subsystems' respective average rotation rates $\langle J_\mathrm{o}\rangle$ and $\langle J_1\rangle$ and chemical driving forces $\Delta\mu_{\mathrm{H}^+}$ and $\Delta\mu_{\mathrm{ATP}}$. This recasts the subsystem efficiency bounds \eqref{eta_y_ineq} and \eqref{eta_x_ineq} as
\begin{subequations}\label{ATP_Synthase_Bounds}
\begin{align}
\etaT\left(1 - \frac{\zeta_1\langle J_1\rangle}{\Delta\mu_\mathrm{ATP}}\right) & \leq\;\;\; \eta_\mathrm{o} && \leq 1 - \frac{\zeta_\mathrm{o}\langle J_\mathrm{o}\rangle}{\Delta\mu_{\mathrm{H}^+}},\\
\etaT\left( 1 - \frac{\zeta_\mathrm{o}\langle J_\mathrm{o}\rangle}{\Delta\mu_{\mathrm{H}^+}}\right)^{-1} & \leq\;\;\; \eta_1 && \leq  \left( 1 - \frac{\zeta_1\langle J_1\rangle}{\Delta\mu_\mathrm{ATP}}\right)^{-1}.
\end{align}
\end{subequations}

The six quantities composing the above bounds can be estimated from experimental data and theoretical calculations. Consider the bovine mitochondria, where many of the relevant quantities have been determined experimentally for ATP synthase far from stall. The chemical driving forces are estimated as $\Delta\mu_\mathrm{ATP} \approx -7.5 \,k_\mathrm{B}T/\mathrm{rad}$ and $\Delta\mu_{\mathrm{H}^+}\approx 8.3\,k_\mathrm{B}T/\mathrm{rad}$~\cite{silverstein2014exploration}. ATP can be synthesized at a rate of up to 440 molecules/second~\cite{matsuno1988estimation}, and $\mathrm{F}_1$ has been observed rotating at speeds of $\langle J_1\rangle\approx 100\,\mathrm{rot/s}$~\cite{etzold1997turnover}. Accordingly, we estimate the rotational flux to be $\langle J_1\rangle\in[100,150]\,\mathrm{rot/s}$. Ref.~\cite{silverstein2014exploration} found $0.65\leq\etaT\leq1$, so we take $\langle J_1\rangle\leq \langle J_\mathrm{o}\rangle\leq 1.4\langle J_1\rangle$. The friction coefficient of the $\gamma$-shaft rotating within the F$_1$ subsystem has been estimated to be of order $1.5\times 10^{-2}\,\mathrm{pN}\cdot\mathrm{nm}\cdot\mathrm{s}/\mathrm{rad}^2$~\cite{buzhynskyy2007rows}. Accordingly, we take $\zeta_1\in[0.01,0.03]\,\mathrm{pN}\cdot\mathrm{nm}\cdot\mathrm{s}/\mathrm{rad}^2$. Calculations of the rotational friction coefficients in Stokes flow~\cite{okazaki2015elasticity} suggest $\zeta_\mathrm{o} \approx \zeta_1/2$, so we take $\zeta_\mathrm{o} \in[0.005,0.015]\,\mathrm{pN}\cdot\mathrm{nm}\cdot\mathrm{s/rad}^2$.

\begin{figure}[h]
\includegraphics[width=\columnwidth]{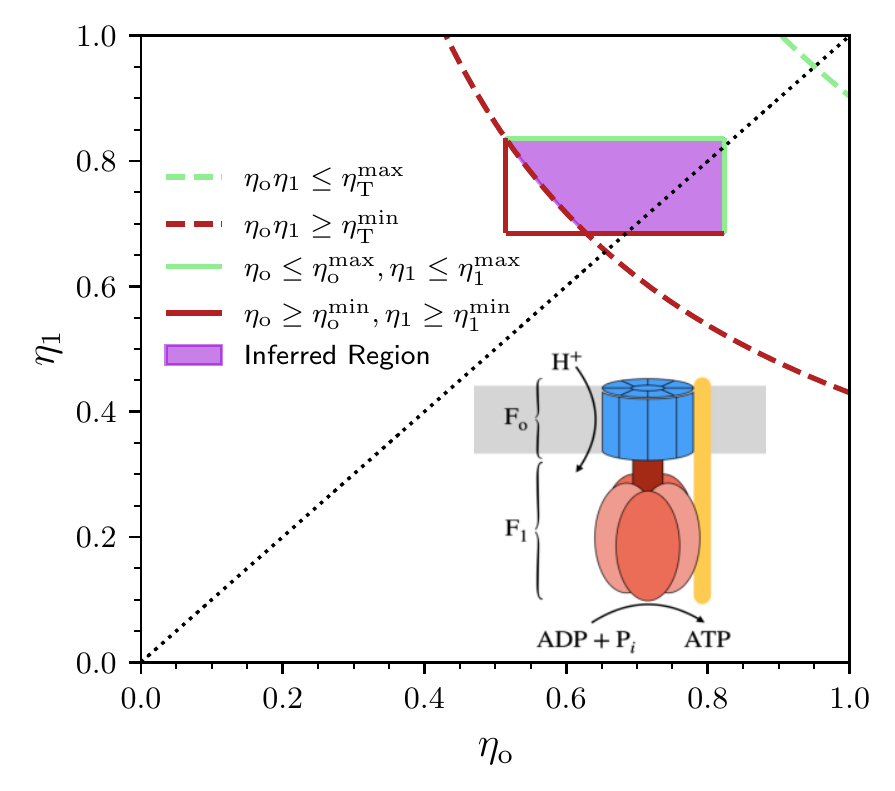} 
\caption{\label{fig:Eta_01_Bounds} Inferred subsystem efficiencies $\eta_\mathrm{o}$ for $\mathrm{F}_\mathrm{o}$ (the $Y$ subsystem) and $\eta_1$ for $\mathrm{F}_1$ (the $X$ subsystem) in $\mathrm{F}_\mathrm{o}\mathrm{F}_1-$ATP synthase. Red horizontal and vertical lines: minimum possible values of the lower bounds~\eqref{ATP_Synthase_Bounds} on $\eta_1$ and $\eta_\mathrm{o}$, respectively, that are consistent with the estimated parameter ranges detailed in the main text. Green horizontal and vertical lines: maximum possible values of the two upper bounds~\eqref{ATP_Synthase_Bounds}. Red and green dashed curves: minimum and maximum possible values of the product $\eta_\mathrm{o}\cdot\eta_1=\etaT$ given our parameter estimates. Purple region: efficiencies satisfying all bounds.}
\end{figure}

Figure~\ref{fig:Eta_01_Bounds} illustrates the joint range of subsystem efficiencies $\eta_\mathrm{o}$ and $\eta_1$ inferred from our bounds \eqref{ATP_Synthase_Bounds}. This significantly constrains the possible subsystem efficiencies within ATP synthase, to $\eta_\mathrm{o}\approx 0.5-0.85$ and $\eta_1\approx 0.7-0.85$. Note that the size and location of the inferred region are somewhat sensitive to the parameter estimates; more precise measurements of physical parameters would allow for tighter thermodynamic inference. Because of the functional form of the Jensen bound, the inferred region is also smaller for higher friction and higher coordinate rates of change. 

\emph{Efficiency of a transport motor pulling a diffusive cargo}.---
Taking $Y$ and $X$ to be the respective one-dimensional positions along a microtubule of a transport motor and its cargo, Eq.~\eqref{finalmotorcargobound} allows estimation of the motor efficiency $\etaM$, quantifying the free-energetic efficiency of a motor pulling against the fluctuating force arising from the motion of the diffusive cargo. Since a coupled motor and cargo have equal average velocity~\cite{leighton2022dynamic}, $\langle\dot{x}\rangle = \langle \dot{y}\rangle = \langle v\rangle$, Eq.~\eqref{finalmotorcargobound} further simplifies to
\begin{equation}\label{simple_motor_bound}
\etaS\leq \eta_\mathrm{M}\leq 1 - \frac{\Dc}{\Dm}\etaS  \ .
\end{equation}
Here the friction coefficients of Eq.~\eqref{finalmotorcargobound} have been replaced with diffusion coefficients (using the fluctuation-dissipation relation~\cite{seifert2012stochastic}) which are more natural for the motor-cargo system.

Estimating $\etaM$ requires measurements of the diffusion coefficients $\Dc$ and $\Dm$, the average velocity $\langle v\rangle$, and the chemical power consumption $\dot{W}_\mathrm{M}$ by the motor. Ref.~\cite{shtridelman2008force} provides experimental measurements of average velocity for single transport motors (motor number inferred from multimodal velocity distributions) pulling vesicles, as a function of the vesicle diameter. Cargo diffusivity $\Dc$ is estimated from the measured diameter using the Stokes-Einstein relation~\cite{einstein1905motion} and reported measurements of temperature and viscosity. We estimate $\Dm$ by fitting the resulting $\langle v\rangle$ as a function of $\Dc$ to the theoretical prediction $\langle v\rangle/v_\mathrm{max} = \left(1 + \Dm/\Dc\right)^{-1}$~\cite{leighton2022performance} with $v_\mathrm{max}=2\,\mu$m/s (the maximum velocity observed in Ref.~\cite{shtridelman2008force}). Finally, we assume the transport motor tightly couples mechanical motion with chemical energy consumption~\cite{schnitzer1997kinesin} so that $\dot{W}_\mathrm{M} = \langle v\rangle \Delta\mu_\mathrm{ATP}/d$, for step size $d=8$~nm and $\Delta\mu_\mathrm{ATP} = 15\, k_\mathrm{B}T$~\cite[Chapters 3 and 4]{milo2015cell}.

\begin{figure}[h]
\includegraphics[width=\columnwidth]{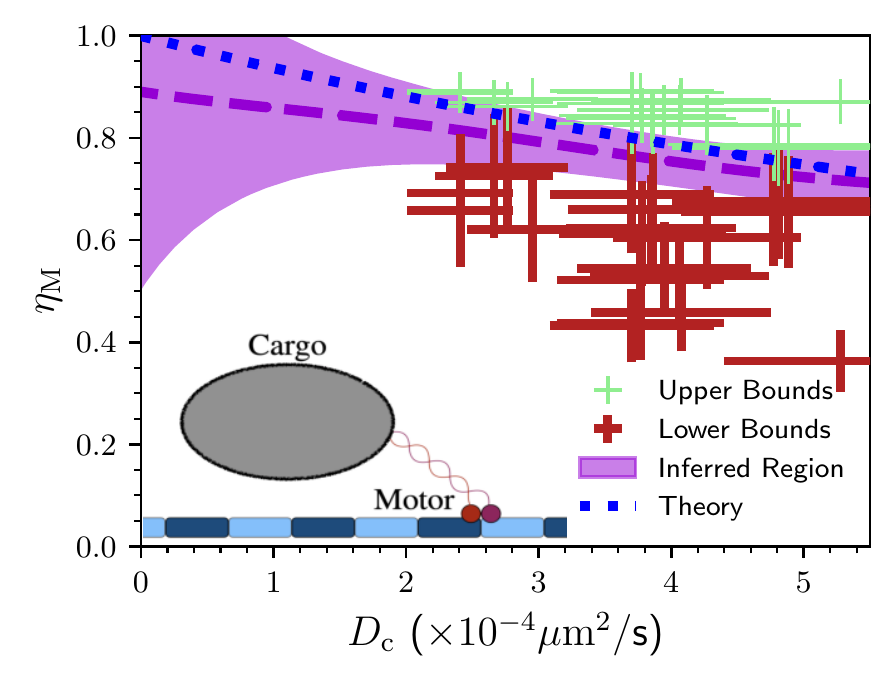}
\caption{\label{fig:Eta_M_Bounds} Upper and lower bounds~\eqref{simple_motor_bound} on the motor efficiency computed from experimental data for a single transport motor (the $Y$ subsystem) pulling a diffusive cargo (the $X$ subsystem)~\cite{shtridelman2008force}. For each paired velocity and cargo diameter from Ref.~\cite{shtridelman2008force}, red and green points respectively denote lower and upper bounds (Eq.~\eqref{simple_motor_bound}) on the motor efficiency, with error bars showing experimental error propagated through Eq.~\eqref{simple_motor_bound}. Purple dashed curve: nonlinear best-fit curve separating the upper and lower bounds computed using a nonlinear support vector machine~\cite{suykens2001nonlinear}. Purple shaded region: corresponding confidence interval. Blue dot-dashed line: theoretical prediction from Ref.~\cite{leighton2022performance}.
}
\end{figure}

Figure~\ref{fig:Eta_M_Bounds} shows the upper and lower bounds on $\etaM$~\eqref{simple_motor_bound} inferred from the above estimates and experimental data. Our method for the first time significantly constrains $\etaM$, suggesting it decreases from $\sim$0.85 to $\sim$0.75 as $\Dc$ increases from $2\,\mu\mathrm{m}^2/\mathrm{s}$ to $5.5\,\mu\mathrm{m}^2/\mathrm{s}$ (corresponding to vesicle diameters from $1.1\,\mu$m $0.4\,\mu$m). Our estimates are consistent with Ref.~\cite{leighton2022performance}'s theoretical prediction $\eta_\mathrm{M}=\left(1 + \Dc/\Dm\right)^{-1}$ (derived assuming the equations of motion~\eqref{LangevinEqs} are linear in $x$ and $y$), which falls entirely within the inferred region of Fig.~\ref{fig:Eta_M_Bounds}.

\emph{Discussion}.---
We derived general lower and upper bounds on the efficiencies of two subsystems composing a bipartite molecular machine in their natural setting, as opposed to in isolation as in typical single-molecule experiments. The measurable quantities required to compute the bounds depend only on the interactions of each molecular machine with its environment; details of the coupling between subsystems need not be understood. Quantifying subsystem efficiencies allows us to determine where free energy is lost in multi-component systems, which will ultimately be critical for the future engineering of synthetic molecular machines. 

The respective efficiencies we inferred for $\mathrm{F}_\mathrm{o}$ and $\mathrm{F}_1$, $\eta_\mathrm{o}\approx 50-85\%$ and $\eta_1\approx 70-85\%$, are somewhat lower than the measured  efficiency of isolated $\mathrm{F}_1$ hydrolyzing ATP, which is nearly $100\%$~\cite{toyabe2010nonequilibrium}. The $\mathrm{F}_\mathrm{o}$ subunit efficiency has likewise been estimated at over $90\%$~\cite{silverstein2014exploration}. Our findings suggest that the $\mathrm{F}_\mathrm{o}$ and $\mathrm{F}_1$ subsystems have different efficiencies acting in concert than when they operate in isolation. One possible reason could be non-tight mechanical coupling between $\mathrm{F}_\mathrm{o}$ and $\mathrm{F}_1$ under physiological conditions; such a ``floppy'' connection imperfectly transfers energy (increasing dissipation) but improves operational speed~\cite{lathouwers2020nonequilibrium} and allows for information flows~\cite{lathouwers2022internal}.

Experimental~\cite{howard1996movement} and theoretical~\cite{lau2007nonequilibrium} investigations of kinesin motors pulling against constant forces suggest motor efficiencies of $40-60\%$. Our inferred range of $70-90\%$ is slightly higher, suggesting that transport motors may attain higher efficiencies when pulling against the variable load produced by a diffusive cargo, their main function within cells. 

It is important to note conceptual differences between previous subsystem efficiencies and those inferred in this Letter. Conventional single-molecule experiments measuring subsystem efficiency (such as Ref.~\cite{toyabe2010nonequilibrium}) typically consider transduction to and from deterministic external reservoirs, hence preclude information flows and are limited to work. Our subsystem efficiencies (Eqs.~\eqref{etaydef} and \eqref{etaxdef}) consider transduction to and from a strongly coupled stochastic subsystem, naturally including information transmission and hence encompassing all free-energy transduction. In contrast with conventional experimental efficiencies that study isolated subsystems in artificial environments, our subsystem efficiencies describe their behaviour in their natural, coupled, \textit{in vivo} context.

Our main results are derived here for $X$ and $Y$ subsystems fully characterized by one-dimensional degrees of freedom, but are more general. The $x$ and $y$ coordinates will in most cases be coarse-grained over many internal degrees of freedom; such a coarse-graining underestimates the true entropy production~\cite{esposito2012stochastic,horowitz2015multipartite}, so our efficiency bounds would loosen but remain valid because they derive from lower bounds on the entropy production rates.

Finally, while we employed the Jensen bound~\eqref{JensenBounds} to derive lower bounds for $\Sigmax$ and $\Sigmay$ and thus derive Eqs.~\eqref{eta_y_ineq} and \eqref{eta_x_ineq}, our framework is far more general. While we know of no other widely applicable sets of lower or upper bounds on $\Sigmax$ and $\Sigmay$ (the well-known thermodynamic uncertainty relation~\cite{gingrich2016dissipation} only restricts the total entropy production rate $\dot{\Sigma}$), any such bounds could be inserted into Eq.~\eqref{eq:TCeqs} to obtain different subsystem efficiency bounds.

\begin{acknowledgments}
\emph{Acknowledgments}.---We thank Shoichi Toyabe (Tohoku Applied Physics) and Jannik Ehrich and Steven Blaber (SFU Physics) for helpful discussions and feedback on the manuscript. This work was supported by Natural Sciences and Engineering Research Council of Canada (NSERC) CGS Masters and Doctoral fellowships (M.P.L.), a BC Graduate Scholarship (M.P.L.), an NSERC Discovery Grant and Discovery Accelerator Supplement (D.A.S.), and a Tier-II Canada Research Chair (D.A.S.).
\end{acknowledgments}

\bibliography{main}

\end{document}